\documentclass[
12pt,
amsmath,
superscriptaddress,
floatfix,
noshowkeyws,
notitlepage,
noshowpacs,
]{revtex4-1}
\usepackage{graphicx}
\usepackage{amsmath}
\usepackage{comment}
\usepackage{amssymb}

\usepackage{subfigure}

\usepackage{color}
\usepackage{soul}

\begin{document}

\title{Underscreening in concentrated electrolytes} 

\author{Alpha A. Lee}
\email{alphalee@g.harvard.edu}
\affiliation{John A. Paulson School of Engineering and Applied Sciences, Harvard University, Cambridge, MA 02138}

\author{Carla Perez-Martinez}
\affiliation{Department of Chemistry, Physical and Theoretical Chemistry Laboratory, University of Oxford, Oxford OX1 3QZ, U.K.}

\author{Alexander M. Smith}
\affiliation{Department of Chemistry, Physical and Theoretical Chemistry Laboratory, University of Oxford, Oxford OX1 3QZ, U.K.}
\affiliation{Department of Inorganic and Analytical Chemistry, University of Geneva, 1205 Geneva, Switzerland} 

\author{Susan Perkin}
\email{susan.perkin@chem.ox.ac.uk} 
\affiliation{Department of Chemistry, Physical and Theoretical Chemistry Laboratory, University of Oxford, Oxford OX1 3QZ, U.K.}

\begin{abstract}

Screening of a surface charge by electrolyte and the resulting interaction energy between charged objects is of fundamental importance in scenarios from bio-molecular interactions to energy storage. The conventional wisdom is that the interaction energy decays exponentially with object separation and the decay length is a decreasing function of ion concentration; the interaction is thus negligible in a concentrated electrolyte. Contrary to this conventional wisdom, we have shown by surface force measurements that the decay length is an increasing function of ion concentration and Bjerrum length for concentrated electrolytes. In this paper we report surface force measurements to test directly the scaling of the screening length with Bjerrum length. Furthermore, we identify a relationship between the concentration dependence of this screening length and empirical measurements of activity coefficient and differential capacitance. The dependence of the screening length on the ion concentration and the Bjerrum length can be explained by a simple scaling conjecture based on the physical intuition that solvent molecules, rather than ions, are charge carriers in a concentrated electrolyte.
\end{abstract}

\makeatother
\maketitle

The structure of electrolytes near a charged surface underpins a plethora of applications, from supercapacitors \cite{conway2013electrochemical} to colloidal self-assembly \cite{evans1999colloidal} and electroactive materials such as ionmonomeric polymer-metal composites \cite{shahinpoor1998ionic}. The structure of dilute electrolytes is relatively well-understood \cite{huckel1923theory,attard1996electrolytes,levin2002electrostatic}. However, dilute electrolytes have a low conductivity because the conductivity is proportional to the concentration of charge carriers. As such, dilute electrolytes are generally unsuitable for many electroactive materials and concentrated electrolytes are preferred up to the point when the viscosity increases significantly. Understanding electrolytes at high concentrations remains a conceptual challenge because the ion-ion Coulomb interaction is strong and long-ranged. The most extreme case of concentrate electrolytes are ionic liquids --- liquids at room temperature which comprise pure ions without any solvent \cite{welton1999room,rogers2003ionic,hallett2011room,fedorov2014ionic}.

To segue into exploring the physics of concentrated electrolytes, we first revisit the physics of dilute electrolytes. The seminal Debye-H\"{u}ckel theory \cite{huckel1923theory} predicts that the interaction between two charged surfaces in an electrolyte decays exponentially with the surface separation \cite{israelachvili2011}. The characteristic decay length, known as the Debye length, is given by 
\begin{equation} 
\lambda_D =\sqrt{\frac{\epsilon k_B T}{4 \pi q^2 c_\mathrm{ion}}} \equiv \frac{1}{ \sqrt{4 \pi l_B  c_\mathrm{ion} }}, 
\label{debyehuckel}
\end{equation}
where $\epsilon$ is the dielectric constant of the medium, $k_B$ the Boltzmann constant, $T$ the temperature, $q$ the ion charge, $ c_\mathrm{ion}$ the ion concentration (which is twice the salt concentration for a 1:1 electrolyte), and 
\begin{equation}
l_B = \frac{q^2 }{\epsilon k_B T}
\end{equation}
is the Bjerrum length. The Bjerrum length is the distance at which the interaction energy between two ions equals the thermal energy unit $k_B T$. The Debye-H\"{u}ckel theory is a mean-field theory valid when $l_B^3  c_\mathrm{ion} \ll 1$, i.e. when the ion-ion separation is far greater than the Bjerrum length and thus the Coulomb interactions can be treated as a perturbation to ideal gas behaviour. Therefore, the Debye-H\"{u}ckel theory is only applicable for dilute electrolytes. 

For concentrated electrolytes, only a handful of analytical results are known. A well-known result pertains to the pair correlation function, $g_{ij}(r)$, which is the probability of finding a particle of component $j$ at a distance $r$ from another
particle of component $i$ \cite{hansen2013theory}. Mathematical analysis of the Ornstein-Zernicke equation reveals that, for particles interacting via a short-ranged \cite{evans1993asymptotic,evans1994asymptotic} or Coulomb \cite{leote1994decay} potential, the asymptotic decay of the correlation function takes the form 
\begin{equation}
 r (g_{ij}(r)-1) \sim A_{ij} e^{-\alpha_0 r} \cos(\alpha_1 r + \theta_{ij}), \; \; \mathrm{as} \; r \rightarrow \infty. 
\label{asym_decay}
\end{equation}
Crucially, Equation (\ref{asym_decay}) implies that all correlation functions in the system decay with the same rate $\alpha_0$ and oscillate with same wavelength $2\pi/\alpha_1$ in the asymptotic limit; only the amplitude $A_{ij}$ and phase $\theta_{ij}$ are species-dependent. In an electrolyte solution, $1/\alpha_0$ is the electrostatic screening length and $2\pi/\alpha_1$ the characteristic correlation wavelength. Equation (\ref{asym_decay}) is a general asymptotic result for the decay of correlations that is independent of the electrolyte model. Analytical expressions for $\alpha_0$ and $\alpha_1$ could be obtained for the restricted primitive model \cite{attard1993asymptotic,leote1994decay}. However, the restricted primitive model does not explicitly account for space-filling solvent molecules and thus may not capture certain important features of screening in electrolytes (c.f. Section \ref{scaling_theory}). Without considering specific models to compute $\alpha_0$ and $\alpha_1$, we will use the Equation (\ref{asym_decay}) to organise our discussion about different theories. 

The decay of correlations in the bulk electrolyte is directly related to the decay of interactions between charged surfaces, measurable via techniques such as the Surface Force Balance (SFB).  To illustrate why this is the case, recall from Equation (\ref{asym_decay}) that the asymptotic wavelength is the same for all correlation functions. Therefore, if we consider two large charged spheres of radius $R$ immersed in the electrolyte, their asymptotic pair correlation function is given by $r (g_{ss}(r)-1) \sim A_{ss} e^{-\alpha_0 r} \cos(\alpha_1 r + \theta_{ss})$. Thus the potential of mean force $ v(r) \sim - k_B T \log g(r) \sim A_{ss} e^{-\alpha_0 r} \cos(\alpha_1 r + \theta_{ss})/r$. As the concentration of the large spheres is negligible compared to the ions and solvent, $\alpha_0$ and $\alpha_1$ are independent of the properties of the spheres. The interactions between the charged plates in the SFB decays in the same way as the same as the interactions between two charged spheres of radius $R\rightarrow \infty$. Therefore, within this picture, the electrostatic screening length and characteristic correlation wavelength measured by SFB is the same as that for the bulk electrolyte.  

We first consider the characteristic correlation wavelength $2\pi/\alpha_1$. The Debye-H\"{u}ckel theory for dilute electrolytes corresponds to the limiting case $\alpha_1 =0$ and $\alpha_0 =1/\lambda_D$. However, a finite oscillatory period emerges ($\alpha_1>0$) when $a/ \lambda_D = \sqrt{4 \pi l_B c_0 a^2} \gtrsim \sqrt{2}$, where $a$ is the ion diameter \cite{attard1993asymptotic,leote1994decay,attard1996electrolytes}. This threshold value of $a/\lambda_D$ is widely known as the Kirkwood line \cite{kirkwood1936statistical}, first reported by John Kirkwood in 1936. In other words, pass the Kirkwood line, the decay of ion-ion correlations switches from a monotonic exponential decay to a damped oscillatory decay. In the context of ionic liquids, the presence of an oscillatory decay of ion charge density away from charged interfaces has been called ``overscreening'' \cite{bazant2011double}. Integral equation theories also predict that the density-density correlation function becomes oscillatory and has a decay length that is longer than the charge-charge correlation function at an even higher electrolyte concentration \cite{attard1993asymptotic,leote1994decay,attard1996electrolytes}; this is termed ``core-dominated'' decay. 

The subject of this paper is the electrostatic screening length $1/\alpha_0$. The Debye-H\"{u}ckel theory (\ref{debyehuckel}) predicts that the electrostatic screening length decreases as the electrolyte concentration increases. Direct experimental measurements of this screening length for concentrated electrolytes is relatively scarce, perhaps a surprise as the theory of electrolyte solutions has received significant attention in the past century \cite{robinson2002electrolyte}. The first sign that the Debye-H\"{u}ckel screening length is qualitatively awry for concentrated electrolytes is a series of SFB studies showing that the interaction force between charged surfaces in an ionic liquid decays exponentially, but with a decay length that is orders of magnitude larger than the Debye length or the ion diameter \cite{gebbie2013ionic, gebbie2015long}. It was then shown, via SFB measurements of the screening length in ionic liquid-solvent mixtures and alkali halide salt solutions, that the long electrostatic screening length is not unique to pure ionic liquids: the electrostatic screening length in concentrated electrolytes \emph{increases} with ion concentration, contrary to the predictions of the Debye-H\"{u}ckel theory \cite{smith2016electrostatic}. Moreover, we provide empirical evidence that the screening length scales as 
\begin{equation}
\lambda_S \sim l_B  c_\mathrm{ion} a^3. 
\label{underscreening}
\end{equation}
In the remainder of this paper, we will term Equation (\ref{underscreening}) ``underscreening''. The electrolyte solution ``underscreens'' charged surfaces in the sense that the interaction between charged surfaces is significantly longer-ranged than the Debye-H\"{u}ckel regime of ions behaving as a weakly interacting gas.  

To allay potential confusion, we emphasize that ``underscreening'' and its cognate ``overscreening'' \cite{bazant2011double} are two distinct parameters in the decay of ion-ion correlation, Equation (\ref{asym_decay}). Underscreening pertains to the anomalously long electrostatic screening length and overscreening pertains to a finite oscillatory period. Therefore, mathematically speaking, overscreening and underscreening could occur together if an electrolyte has an oscillatory decay of ion-ion correlation with a decay length that follows the scaling (\ref{underscreening}). However, experimentally oscillations are measured only in the near-surface region and no oscillatory component is detected in the long-ranged component of the surface force \cite{gebbie2013ionic, gebbie2015long,smith2016electrostatic}. 

In this paper, we first discuss the experimental evidence for underscreening and report a new set of experiments verifying the scaling relationship (\ref{underscreening}). We then show how underscreening is reflected in two classic properties of electrolytes: the activity coefficient and differential capacitance. Finally, we propose a scaling conjecture to understand the phenomenology of underscreening. 

\section{Experimental measurements of the screening length}
\label{Expt_measurements}

The screening lengths, $\lambda_S$, of electrolyte solutions were determined from direct measurements of the change in the interaction force with distance between two charged mica plates across the electrolyte. The apparatus used for such measurements, called the surface force balance (SFB; see Figure~\ref{Fig:ExperimentalFigureOne}), employs white light interferometry to determine the separation between the plates to $\sim$ 0.1 nm. The mica plates are supported on cylindrical lenses (each of radius $\sim$ 1 cm) and mounted in crossed-cylinder configuration; the arrangement is geometrically equivalent to a sphere of radius  1 cm approaching a flat plate. The symmetry and well-defined geometry make the setup particularly useful for quantitative comparison to theory; the technique has been used over the past few decades to study forces across dilute electrolytes \cite{Israelachvili1978}, simple molecular liquids \cite{Horn1981}, and soft matter \cite{israelachvili2011}. In the case of dilute electrolytes the surface force is dominated by a repulsive osmotic pressure, increasing exponentially as $D$ decreases, with decay length equal to $\lambda_D$ which decreases with increasing concentration in accordance with Equation (\ref{debyehuckel}).

In contrast to the measurements in dilute electrolytes, surface force measurements across pure ionic liquids have revealed short-range oscillations reminiscent of structural forces in  molecular liquids \cite{Perkin2010} and, beyond the oscillatory region, monotonic screening extending to distances far greater than predicted by simple application of Debye-H\"{u}ckel theory \cite{gebbie2013ionic}. In a study aimed at connecting up the dilute electrolyte and ionic liquid ends of the electrolyte spectrum, some of us recently reported a non-monotonic trend in the asymptotic screening length with concentration \cite{smith2016electrostatic}. In this section we describe those experiments in detail and investigate the scaling behaviour of $\lambda_S$ with $c_\mathrm{ion}$ and, separately, the scaling of $\lambda_S$ with $l_B$ achieved by varying solvent $\epsilon$ at constant $c_\mathrm{ion}$. 

\subsection{Experimental details}

\begin{figure}
\centering
\includegraphics{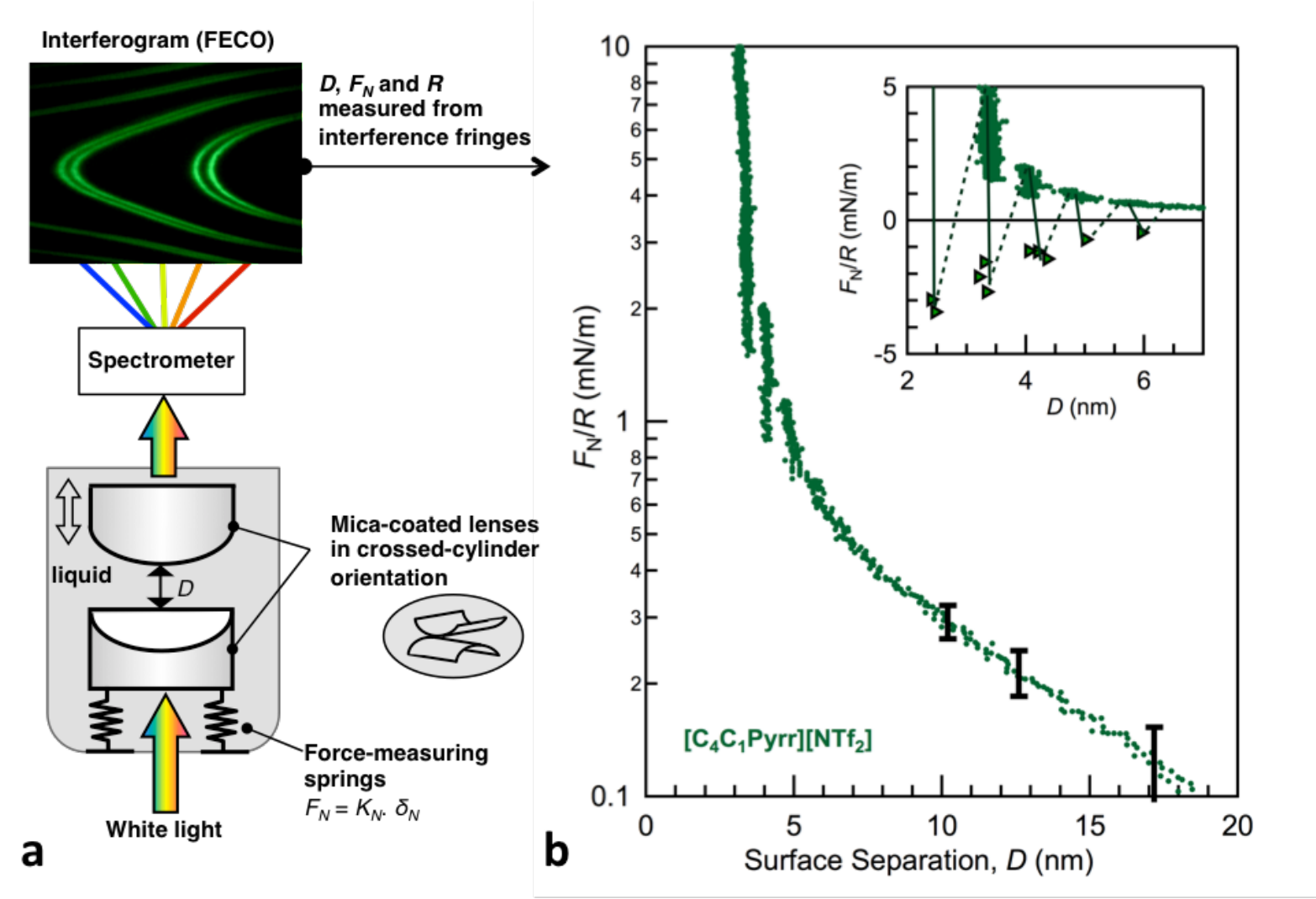} 
\caption{Schematic diagram (a) showing the essential features of the SFB experiment. White light is passed through an interferometric cavity comprising two hemi-cylindrical lenses in crossed-cylinder orientation. Mica sheets (not shown) are back-silvered to allow for white light reflection before mounting on the lenses. The mica sheets are immersed in the liquid electrolyte of interest. Analysis of the fringes of equal chromatic order (FECO) are analysed during approach of the cylindrical lenses in order to calculate the mica-mica separation distance ($D$), force ($F_N$), and effective radii of curvature of the lenses ($R$). An example of a single measurement is shown in (b) for the pure ionic liquid [C$_4$C$_1$Pyrr][NTf$_2$]; the data are shown on log-linear plot to demonstrate the exponential decay at longer range and, in the inset, on a linear-linear plot to reveal the oscillatory region with negative force minima at small $D$.  Replotted from Ref.~\cite{smith2016electrostatic}.}
\label{Fig:ExperimentalFigureOne}
\end{figure}

In the SFB experiments, white light interferometry is used to determine the forces between two molecularly smooth mica surfaces separated by a thin film of electrolyte (see Figure~\ref{Fig:ExperimentalFigureOne}) \cite{Klein1998}. The mica sheets of equal thickness are backsilvered to create a partially reflecting and partially transmitting mirror before gluing onto the lenses and injection of liquid in the gap between the mica surfaces. The resulting silver-mica-liquid-mica-silver stack acts as an interferometric cavity. Bright columnated white light incident on interferometer, dispersed with a spectrometer, emerges as a set of bright fringes of equal chromatic order (FECO). 

The bottom lens is mounted on a horizontal leaf spring, while the top lens is mounted on a piezo-electric tube (PZT). By expanding the PZT, the top surface is brought at constant velocity towards the bottom surface from separations $D$ of 200-400 nm to $D$ of one or a few molecular diameters. The rate of approach is sufficiently slow that there is no measurable hydrodynamic contribution to the force, as evidenced by the insensitivity of the measured forces to small changes of rate of approach.  The FECO pattern is captured by a camera at rates of approximately 10 frames per second. At large separations, there is no normal force on the spring, but as the lenses are brought in to contact, the normal forces arising from interactions between the surfaces will cause bending of the spring. The deflection of the spring, and thus the normal force, can be inferred from the interferometric pattern. The force between the surfaces can then be related to the interaction energy $E$ using the Derjaguin approximation \cite{israelachvili2011}. Therefore, $E=F/2\pi R$, where $R$ is the local radius of curvature between the lenses, of the order of 1 cm for these experiments.

In order to vary solvent $\epsilon$ at constant $c_\mathrm{ion}$ we used solutions of ionic liquid at fixed 2M concentration in solvents of varying polarity. The ionic liquid was 1-butyl-1-methypyrrolydinium bis[(trifluoromethyl)sulfonyl]imide (abbreviated [C$_4$C$_1$Pyrr][NTf$_2$], Iolitec 99.5 \%), and the molecular solvents were propylene carbonate (Sigma-Aldrich, anhydrous 99.7 \%), dimethyl sulfoxide (Sigma-Aldrich, anhydrous 99.9\%), acetonitrile (Sigma Aldrich, anhydrous 99.8\%), benzonitrile (Sigma-Aldrich, anhydrous 99\%) and butyronitrile (Fluka, purity $\geq$99\%).

The FECO fringes were analysed using the method outlined by Israelachvilii \cite{Israelachvili1973}; our analysis uses the refractive index values of the bulk mixture to compute the separation between the mica surfaces. The refractive index of the mixtures of 2M [C$_4$C$_1$Pyrr][NTf$_2$] in dimethyl sulfoxide and in benzonitrile were measured to be 1.441 and 1.461, respectively, using an Abbe 60 refractometer; for mixtures of 2M [C$_4$C$_1$Pyrr][NTf$_2$] in propylene carbonate, butyronitrile, and acetonitrile, the FECO analysis used estimated refractive index values of 1.422, 1.408, and 1.380, respectively. These estimated values are weighted average values between the refractive index of the pure ionic liquid (1.425; measured by supplier), and the refractive index of the solvents (1.4189 for propylene carbonate and 1.3842 for butyronitrile, both from CRC handbook \cite{lide2003crc}, and 1.344 for acetonitrile (provided by supplier). Calculation of such a weighted average for dimethyl sulfoxide and benzonitrile solutions led to values in very good agreement with our direct measurements.  

Several precautions are taken to ensure the purity and stability of the liquid mixtures during the measurements. In all experiments, the ionic liquid [C$_4$C$_1$Pyrr][NTf$_2$] was dried in vacuo (10$^{-2}$ mbar, 70$^{\circ}$C) for several hours to remove residual water. In the case of acetonitrile, butyronitrile and propylene carbonate experiments, the liquid was obtained from freshly opened bottles, while for benzonitrile measurements and some of the dimethyl sulfoxide experiments, the bottles had been opened within two weeks of the measurement. The dried ionic liquid was then mixed with the solvents and introduced in between the lenses within a few minutes, in order to minimise exposure of the mixture to atmospheric moisture. In all cases, the liquid film between the mica surfaces was in contact with a large bulk reservoir. For solutions of ionic liquid with dimethyl sulfoxide, benzonitrile, and propylene carbonate, a droplet of solution of approximately 20 $\mu$L was injected between the lenses. In the case of acetonitrile and butyronitrile solutions, which are volatile, the bottom lens was immersed in a bath of the solution, and for acetonitrile tests, additional solvent was introduced in the SFB chamber to create a saturated solvent vapour and thus minimise evaporation. The drying agent P$_2$O$_5$ was also introduced in the chamber to capture any residual water vapour. Different glues were used to attach the mica sheets to the lenses, depending on the compatibility of the solvents: glucose (Sigma-Aldrich, 99.5\%) was used as glue for propylene carbonate experiments, EPON 1004 (Shell Chemicals) was used for benzonitrile, acetonitrile and butyronitrile soltuions, and paraffin (Aldrich, melting point 53-57$^{\circ}$) was used for dimethyl sulfoxide experiments.

\subsection{Experimental measurements varying $c_{\textrm{ion}}$}

An example of the measured interaction force between two mica plates as a function of separation, $D$, across a pure ionic liquid, [C$_4$C$_1$Pyrr][NTf$_2$], is shown in Figure~\ref{Fig:ExperimentalFigureOne}(b). As the surfaces approach from large $D$ they experience a repulsive force, exponentially increasing with decreasing $D$, eventually giving way to an oscillatory region at $D \lesssim 5-8~\mathrm{nm}$. The key signature of oscillatory forces is the presence of minima in the profile as detected on retraction of the surfaces; these are shown using a linear scale in the inset to Figure~\ref{Fig:ExperimentalFigureOne}(b). Interpretation of the structural features in ionic liquids leading to such oscillatory forces has been discussed in the past \cite{Perkin2011,fedorov2014ionic}, the details of this near-surface region depend on ionic liquid molecular features such as cation-anion size asymmetry and ion amphiphilicity, surface chemistry and surface charge. Here we focus instead on the monotonic tail of the interaction force---in this pure ionic liquid, the tail is measurable above our resolution limit from about 20 nm (several tens of ion diameters)---which we will show to be relatively insensitive to the molecular features of the ionic liquid or electrolyte. The exponential decay length in the asymptotic limit is taken as the screening length $\lambda_S$. 

\begin{figure}
\centering
\includegraphics{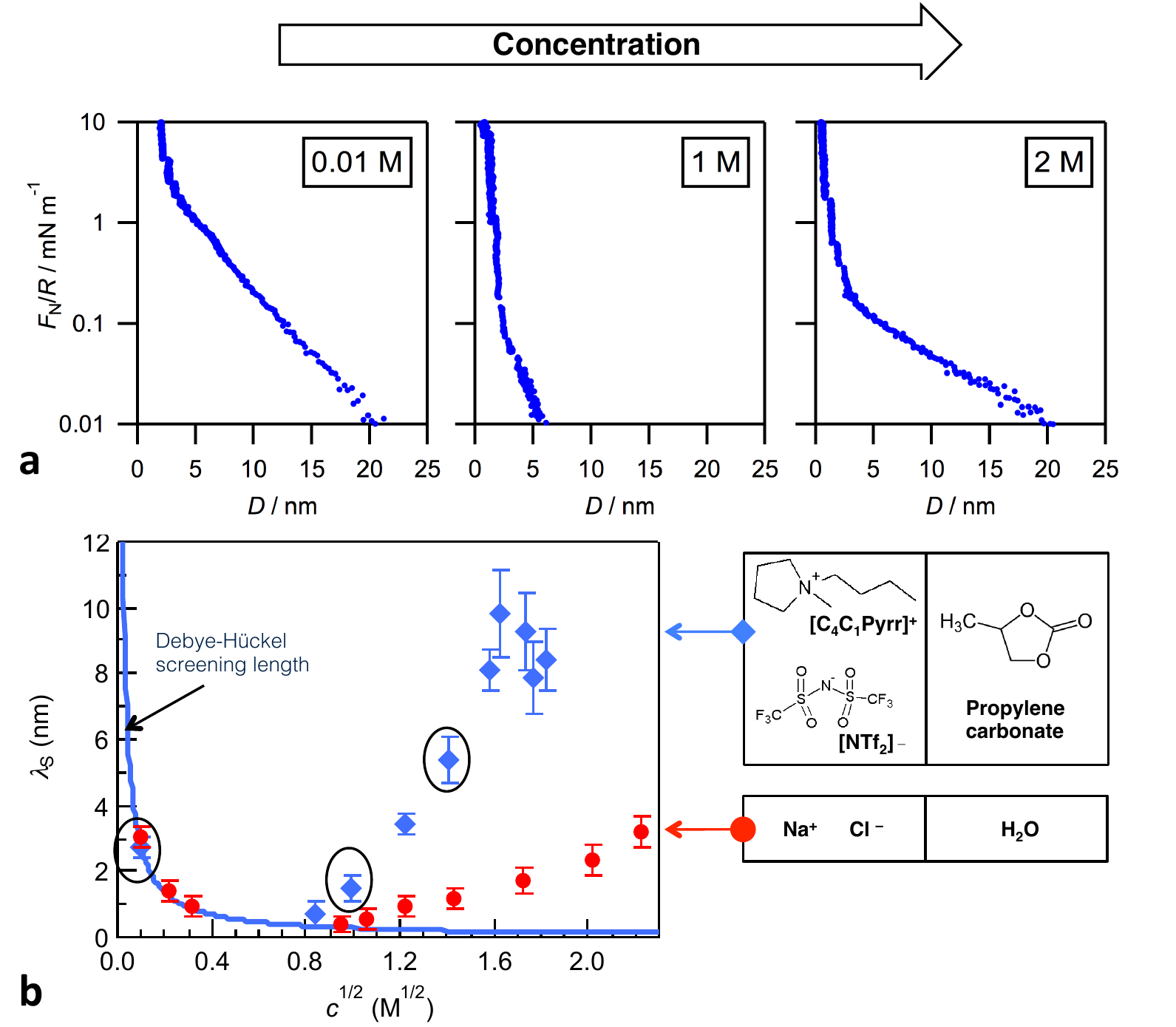} 
\caption{(a) Example measurements of the normalised force between mica sheets as a function of separation, $D$, across mixtures of [C$_4$C$_1$Pyrr][NTf$_2$] and propylene carbonate at $c$ = 0.01 M, 1.0 M and 2.0 M. The concentrations are chosen to demonstrate the non-monotonic variation in the long-range decay with concentration. (b) Screening length of the long-range (asymptotic) component of the surface force plotted as a function of $c^{1/2}$, and for two different electrolytes: [C$_4$C$_1$Pyrr][NTf$_2$] in propylene carbonate, and NaCl in water. Data in (b) are replotted from Ref.~\cite{smith2016electrostatic}. The circled data points arise from the three force profiles in (a).}
\label{Fig:ExperimentalFigureTwo}
\end{figure}

We studied the variation of $\lambda_S$ with $c_{\textrm{ion}}$ in a mixture of [C$_4$C$_1$Pyrr][NTf$_2$] with propylene carbonate (molecular solvent), chosen for their miscibility and liquidity over the full range of mole fraction, from pure solvent to pure salt, at room temperature. Figure~\ref{Fig:ExperimentalFigureTwo}(a) shows three force profiles chosen at concentration points to demonstrate the clear decrease in $\lambda_S$ between 0.01 M ($\lambda_S=2.7\pm 0.3 \; \mathrm{nm}$) and 1.0 M ($\lambda_S=1.05\pm 0.4 \; \mathrm{nm}$), and the subsequent increase in $\lambda_S$ between 1.0 M and 2.0 M ($\lambda_S=5.4\pm 0.7 \; \mathrm{nm}$). Figure~\ref{Fig:ExperimentalFigureTwo}(b) shows how $\lambda_S$ varies with $c_{\textrm{ion}}^{1/2}$, and also shows similar measurements made for NaCl in water. It is clear that in both cases there exists a minimum in $\lambda_S$ at intermediate concentration.

\begin{figure}
\centering
\includegraphics[scale=1.2]{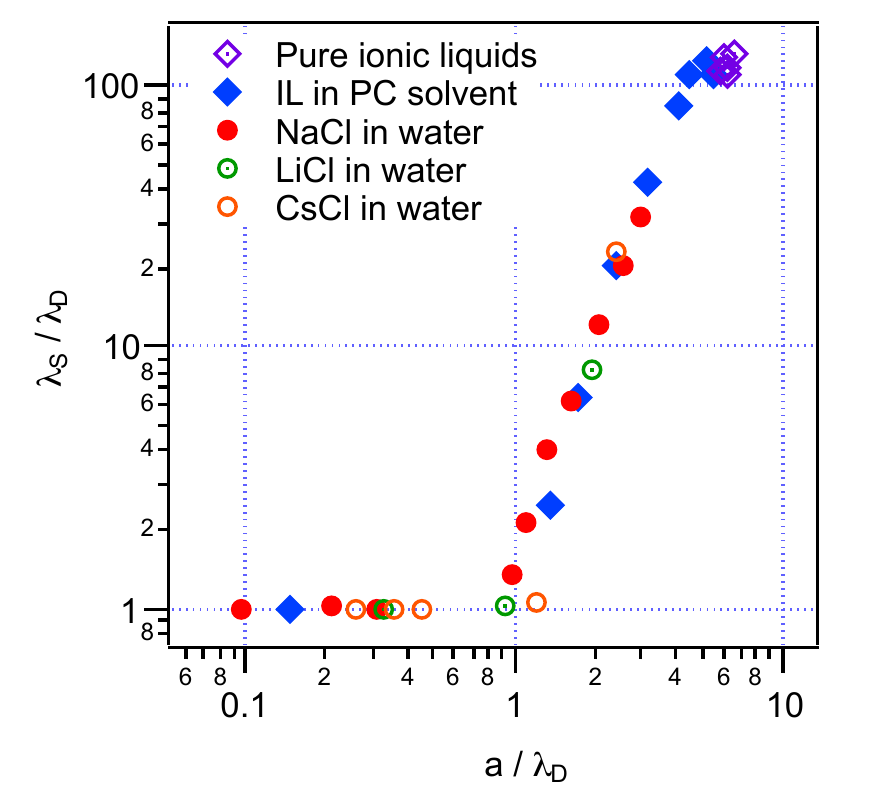} 
\caption{Experimentally measured screening length, $\lambda_S$, normalised by $\lambda_D$, plotted against $a/\lambda_D$ for a range of pure ionic liquids, ionic liquid mixed with propylene carbonate molecular solvent, and various 1:1 inorganic salts in water. Data for LiCl and CsCl are literature values \cite{Baimpos2014,Pashley1984}.}
\label{Fig:ExperimentalFigureThree}
\end{figure}

The realisation that NaCl in water at sufficiently high concentration shows the same divergence of screening length as observed in ionic liquids led us to hypothesise that the origin of the anomalous $\lambda_S$ lies in electrostatic interactions between ions, rather than in a mechanism dependent on chemical features such as hydrogen bonding or nanoscale aggregation of non-polar domains. This indeed appears to be the case, as demonstrated by the collapse of all data points when the screening length is scaled by the Debye length and the concentration is scaled by the dielectric constant and ion diameter, as shown in Figure~\ref{Fig:ExperimentalFigureThree}. We note that the dielectric constant varies substantially as a function of ion concentration; the dielectric constants of ionic liquid solutions are calculated using effective medium theory \cite{Bergman1978}, and the dielectric constant of alkali halide solutions are taken from the literature \cite{buchner1999dielectric,gavish2016dependence}. Included in Figure \ref{Fig:ExperimentalFigureThree} are also a wide range of pure ionic liquids, and some further 1:1 inorganic salts in water; the common scaling appears to be general across these electrolytes. The abscissa in Figure~\ref{Fig:ExperimentalFigureThree} is the nondimensional quantity $a/\lambda_D$, where $a$ is the mean ion diameter in the electrolyte; the ion diameter of ionic liquid is estimated from X-ray scattering experiments \cite{santos2011communication}, and we take the ion diameter of alkali halide salts to be the unhydrated ion diameter \cite{smith2016electrostatic}.  $a/\lambda_D$ scales as $(c_{\textrm{ion}}/\epsilon)^{1/2}$. As we will show in the following section, $a/\lambda_D$ is also an important parameter describing the scaling of the chemical potential and activity in electrolyte solutions. 

There are two distinct scaling regimes in Figure \ref{Fig:ExperimentalFigureThree}, which we call ``low" and ``high" concentration. At low concentration, where $a/\lambda_D<1$, the measured screening length is the Debye length i.e. $\lambda_S/\lambda_D=1$. This persists until the point at which the Debye length shrinks to the ion diameter, $a=\lambda_D$. At high concentration, when $a/\lambda_D>1$, the scaling switches to a power law:

\begin{equation} 
\frac{\lambda_S}{\lambda_D}\sim \left(\frac{a}{\lambda_D}\right)^3
\end{equation}
which is equivalent to $\lambda_S\sim  c_{\textrm{ion}} a^3 l_B$. That is to say, our measurements suggest that in the high-concentration regime the screening length scales linearly with Bjerrum length.

\subsection{Experimental measurements varying $l_B$}

\begin{figure}
\centering
\includegraphics[scale=1]{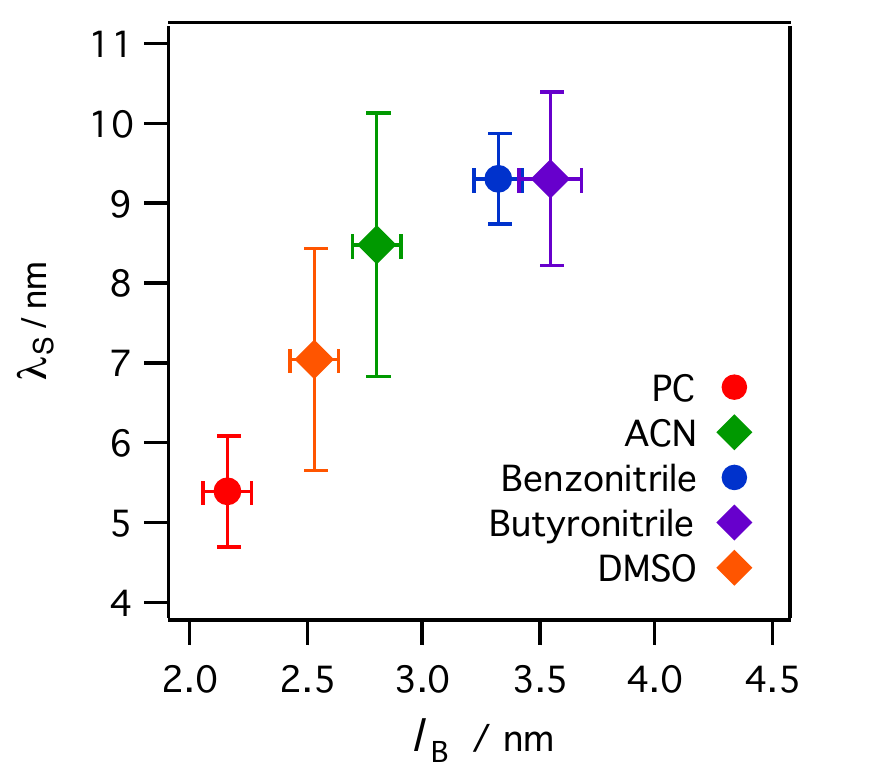} 
\caption{Testing the scaling of measured screening length, $\lambda_S$, with $l_B$. Each data point corresponds to a 2M solution of [C$_4$C$_1$Pyrr] [NTf$_2$] in a different solvent -- and therefore different dielectric constant -- thereby $l_B$ is varied at fixed $c_{\textrm{ion}}$. The solvents are as shown in the legend and their dielectric constants range from 64.0 (propylene carbonate) to 22.8$\pm 2.1\ $(butyronitrile) for the pure solvents. Dielectric constants for the 2M solutions are calculated using Effective Medium Theory. The vertical error bars arise from scatter between the experimental decay length measured in different experiments and different force profiles in the same experiment. The horizontal error bars arise from uncertainty in the dielectric constants of ionic liquid and solvent.}
\label{Fig:ExperimentalFigureFour}
\end{figure}

The experiments described above consist of a survey across the concentration spectrum for two classes of electrolyte. This series of experiments, however, does not provide the most direct test of the scaling of screening length because along this axis $c_{\textrm{ion}}$ and $\epsilon$ are coupled: each increment in concentration also leads to an alteration of dielectric constant of the mixture. Therefore in order to test the apparent linear relationship between $\lambda_S$ and $l_B$ in the high concentration regime we next carried out a series of experiments where the Bjerrum length was varied at fixed (high) salt concentration. This was achieved using a range of molecular solvents with a wide range of dielectric constants mixed with salt at fixed $c_{\textrm{ion}}$ ($2M$). We measured the force between mica plates across each of these electrolytes in the SFB: in each case a long-ranged exponential decaying force was apparent, qualitatively similar to those described above, and the asymptotic decay length was extracted. The resulting screening lengths are plotted against the Bjerrum length of the electrolyte mixture in Figure~\ref{Fig:ExperimentalFigureFour}. Considerable experimental error arises from these measurements employing volatile solvents (particlarly ACN) and when removing traces of water is difficult (for DMSO). Nonetheless it is clear that the screening length increases with Bjerrum length, and the data are  consistent with a linear scaling of $\lambda_S$ with $l_B$ as predicted by the underscreening relationship (Equation (\ref{underscreening})). 

\begin{figure}
\centering
\subfigure{\includegraphics[scale=0.4]{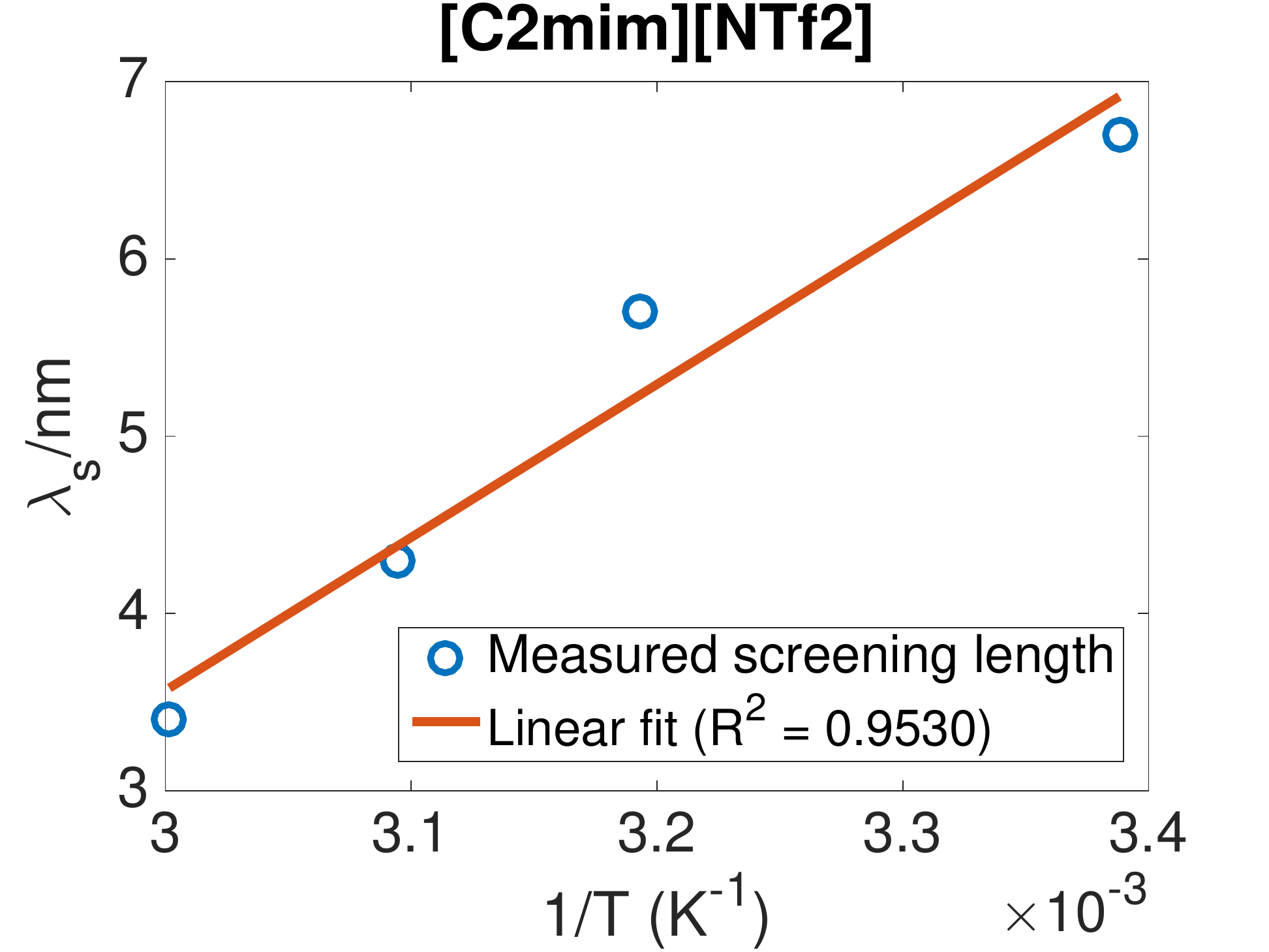}}
\subfigure{\includegraphics[scale=0.4]{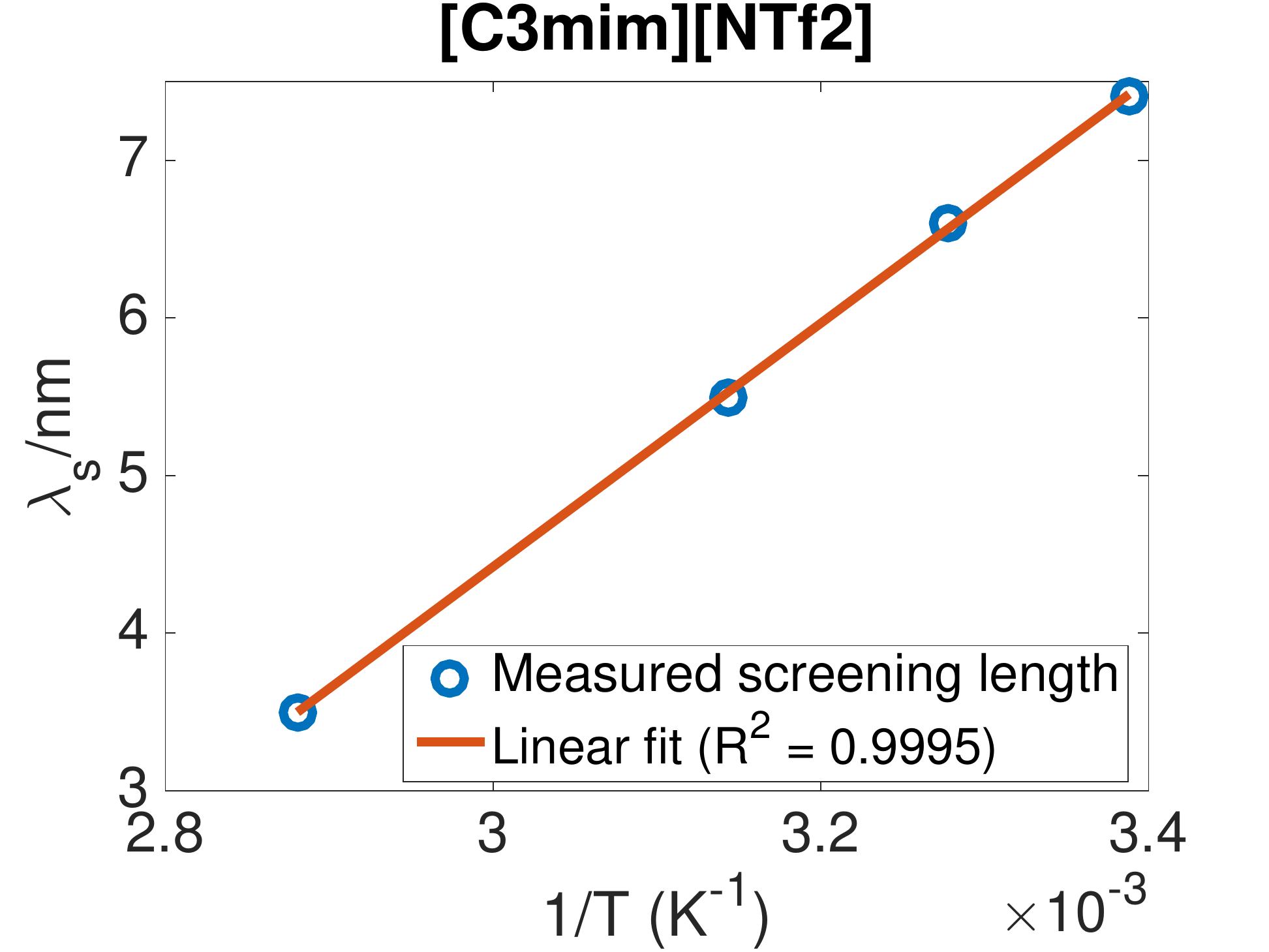}}
\subfigure{\includegraphics[scale=0.4]{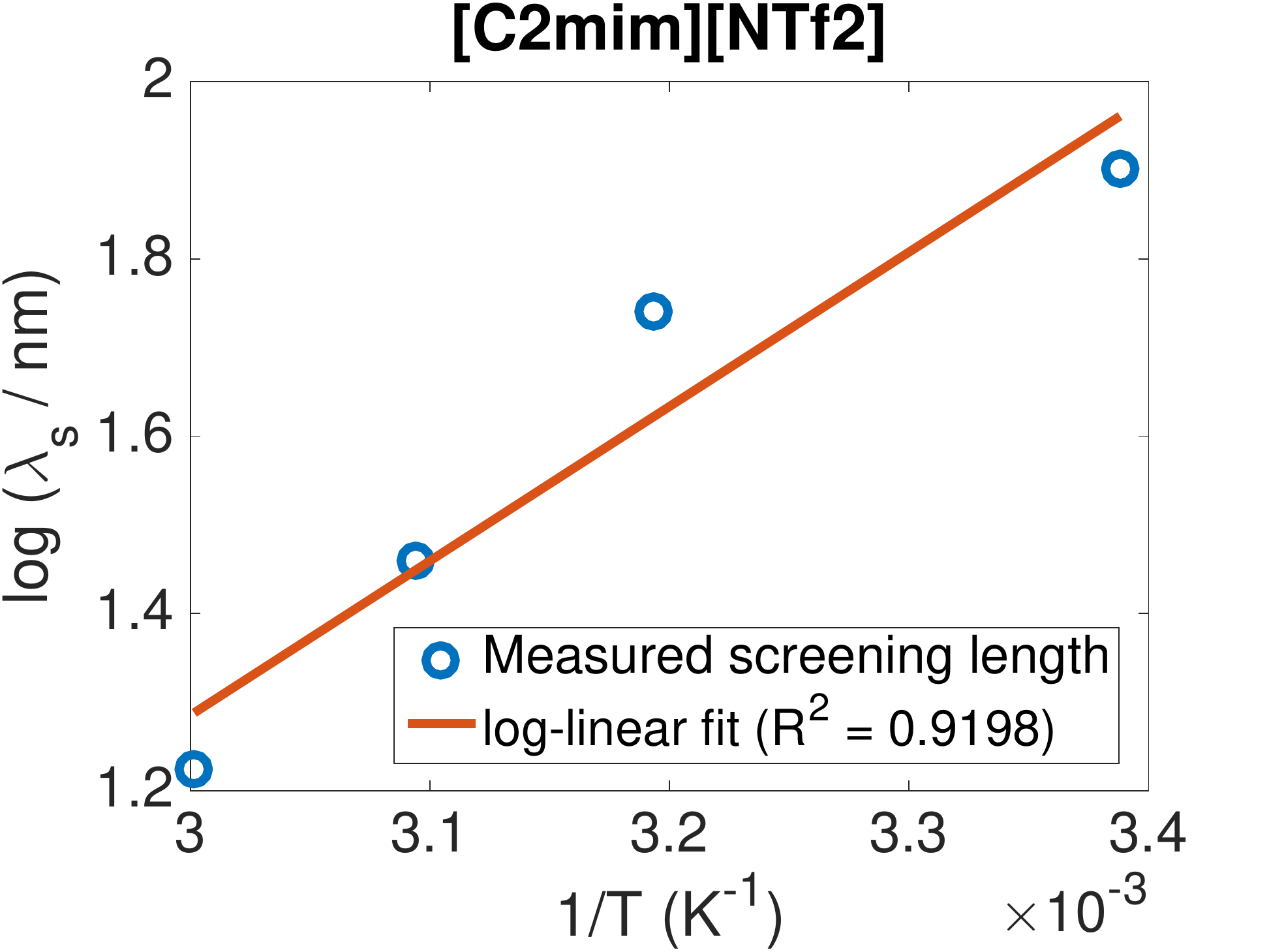}}
\subfigure{\includegraphics[scale=0.4]{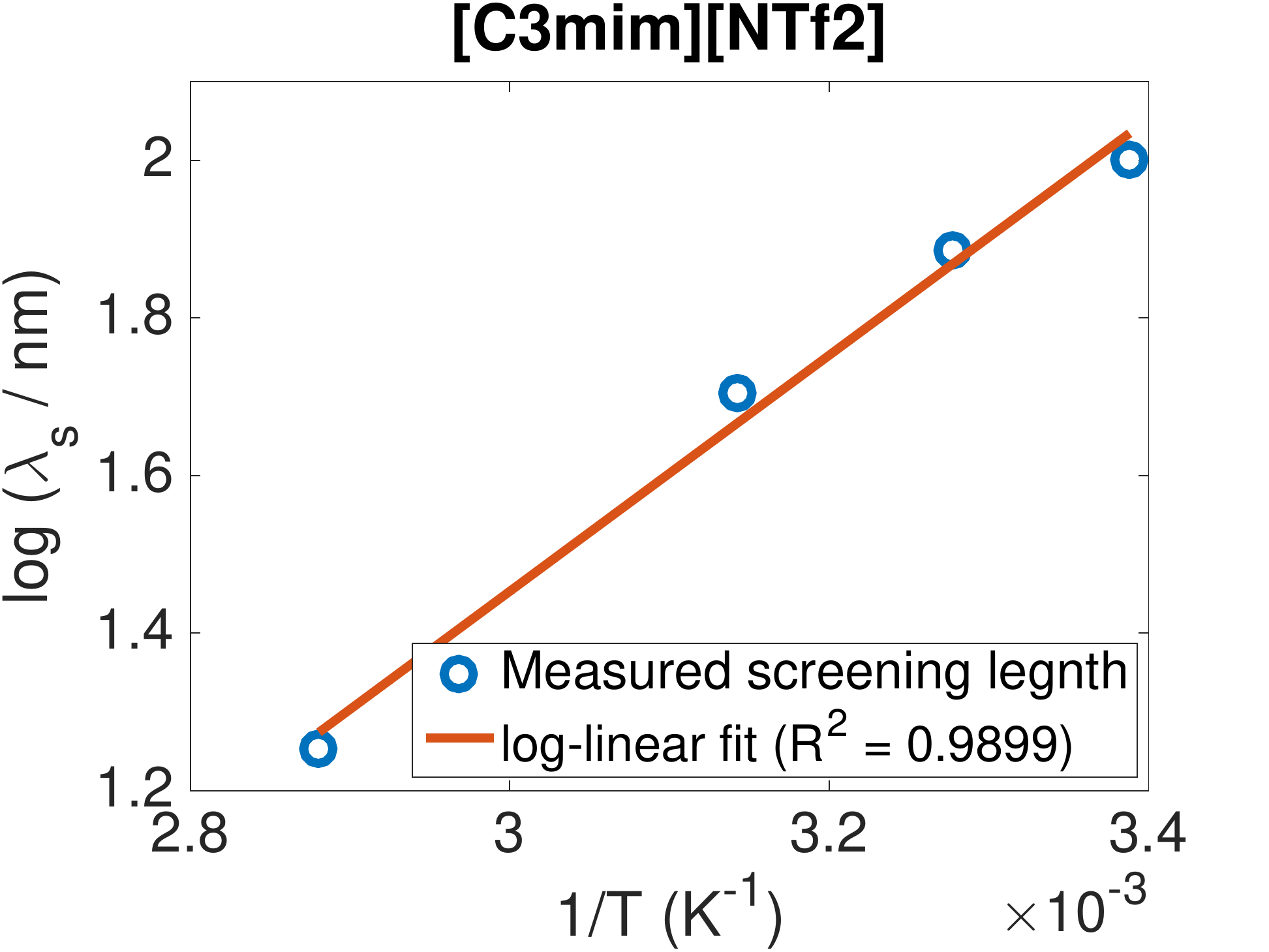}}
\caption{The hypothesis that $ \lambda_S \propto 1/T$ is better supported by data than  $\log( \lambda_S) \propto 1/T$. The screening lengths are taken from ref \cite{gebbie2015long}.} 
\label{gebbie2015}
\end{figure}

Finally, we note that the proposed underscreening scaling of $\lambda_S$ with $l_B$ also implies that $\lambda_S \sim 1/T$, and therefore experiments with varying temperature also provide a test of the underscreening relationship. In a recent paper comparing the screening lengths in $\mathrm{[C_2mim][NTf_2]}$ and $\mathrm{[C_3mim][NTf_2]}$ at different temperatures it was indeed found that the screening length decreases with increasing temperature \cite{gebbie2015long}. However the activated mechanism proposed there led to the suggestion of an Arrhenius dependence on temperature, i.e. $\log( \lambda_S) \sim 1/T$, which is not consistent with the underscreening scaling (Equation (\ref{underscreening})). As such we now revisit the data presented in ref \cite{gebbie2015long}. Figure \ref{gebbie2015}(a)-(b) show the measured screening lengths $ \lambda_S$ vs. $1/T$ and Figure \ref{gebbie2015}(c)-(d) show $\log(\lambda_S)$ vs. $1/T$. The goodness-of-fit of $\log \lambda_S$ against $1/T$ -- the test of Arrhenius dependence -- is actually inferior to that of $ \lambda_S$ against $1/T$, although the difference is slight. Therefore the data in Figure \ref{gebbie2015} are consistent with the scaling $\lambda_S \sim l_B c_0 a^3$. Further studies of temperature dependence of $\lambda_S$ in different electrolytes will help distinguish between the Arrhenius dependence on temperature and the $\lambda_S \sim 1/T$ implied by the underscreening scaling. We note that our analysis and that in ref \cite{gebbie2015long} ignore the dependence of the dielectric constant on temperature; the validity of this assumption must be addressed in future works. 

\section{Relating underscreening to physical properties}

Setting aside the question of why the screening length in concentrated electrolytes is anomalously long, in this section we instead assess whether this anomalously long screening length can be connected to other independently measured physical properties of the electrolytes. We will consider two archetypical properties of concentrated electrolytes -- the activity coefficient of ions and the differential capacitance at the point of zero charge. Analytical expressions that connect the Debye length to both quantities for dilute electrolytes are well-known. An intuitive attempt is to replace the Debye length with the experimentally measured screening length and compare with experimental measured capacitance and activity coefficient. However, one is not sure whether this intuitive approach is consistent. Instead, we will construct a semi-phenomenological free energy functional and derive the relationship between screening length, activity coefficient, and differential capacitance using this free energy. 

Consider a simple Landau-Ginzburg expansion for the free energy $F$ of the electrolyte in response to an infinitesimally small external potential $\delta V(\mathbf{r}) $ and fixed charge distribution $\sigma(\mathbf{r})$. We expand the free energy as a functional of local charge density $\rho = c_+ - c_-$,  
\begin{equation}
F[\rho] =   \frac{e^2}{2\epsilon}\int \int \frac{\left[q \rho( \mathbf{r})+\sigma(\mathbf{r})\right] \left[q \rho (\mathbf{r'})+\sigma(\mathbf{r}')\right]}{| \mathbf{r}-\mathbf{r'}|} \mathrm{d}\mathbf{r}  \mathrm{d}\mathbf{r'} + \int \left[ \frac{p}{2} \rho(\mathbf{r})^2 - \rho(\mathbf{r}) \delta V(\mathbf{r}) \right] \mathrm{d}\mathbf{r},  
\label{free_energy}
\end{equation} 
where $p$ is a phenomenological coefficient. This expansion is valid for infinitesimal charge fluctuations, $\rho \ll c $ where $c = c_+ + c_-$. In this limit, we can assume that the fluctuations in the charge density are independent of the total ion density, and the total ion density is uniform in space. The first term in Equation (\ref{free_energy}) captures the electrostatic interactions between ions and fixed charges, the second term is a local energetic penalty to accumulating charge density. In the Debye-H\"{u}ckel formalism, this term can be derived by linearising the ideal gas entropy, yielding $p = k_B T/c$. However, we will leave $p$ to be a phenomenological parameter which may depend on $c$; we will determine $p$ later by fitting to the experimentally measured screening length. Finally, the last term in Equation (\ref{free_energy}) describes the interaction between the external potential and the induced charge density in the electrolyte. 

We first consider the case with no fixed charge ($\sigma(\mathbf{r})=0$). Minimising the free energy (\ref{free_energy}) yields the Euler-Lagrange equation 
\begin{equation}
p \rho(\mathbf{r}) + \frac{q^2}{\epsilon} \int  \frac{\rho(\mathbf{r'})}{|\mathbf{r} - \mathbf{r'}|} \; \mathrm{d} \mathbf{r'} = \delta V(\mathbf{r}).  
\label{EL-eqn}
\end{equation} 
After performing a three-dimensional Fourier transform on Equation (\ref{EL-eqn}), we arrive at 
\begin{equation}
\hat{\rho}(k) = \frac{ \hat{\delta V}(k)}{p + \frac{4 \pi q^2}{\epsilon k^2}}, 
\end{equation} 
thus the susceptibility is given by $\chi(k) = (p + 4 \pi q^2/(\epsilon k^2))^{-1}$. Therefore, electric field perturbations decay exponentially in the medium with a characteristic screening length
\begin{equation}
\lambda_S =  \sqrt{\frac{ \epsilon p}{ 4 \pi  q^2}}. 
\end{equation}
As such, $p$ could be inferred by measuring the screening length experimentally. A long screening length corresponds to a large value of $p$. 

\subsection{Activity coefficient} 
To motivate the concept of an activity coefficient, we note that the chemical potential of an ideal solution as a function of concentration $c$ reads
\begin{equation}
\mu^{\mathrm{id}} =  \mu^{*} +  k_B T \log c, 
\end{equation}
where $\mu^{*}$ is the standard chemical potential, i.e. the chemical potential of a $1M$ solution at standard conditions. Electrolyte solutions are non-ideal due to ion-ion interactions. The actual chemical potential of the cation/anion can be written as a sum of the ideal solution part and the excess part
\begin{equation}
\mu_{\pm} = \mu^{*} + k_B T \log c_\pm + \mu^{\mathrm ex}_\pm = \mu^{*} +  k_B T \log (\gamma_\pm c_\pm)
\end{equation} 
where $\gamma_\pm = e^{\mu^{\mathrm ex}_\pm/(k_B T)} $ is called the activity coefficient. In other words, the activity, $\gamma_\pm c_\pm$, is a measure of the ``effective concentration'' of species in the system and the activity coefficient, $\gamma_\pm$, is a measure of the deviation of the electrolyte from ideality. The activity coefficient/excess chemical potential is a quantity that has been measured extensively in the literature because of its relevance to electrochemistry \cite{robinson2002electrolyte}.  

To derive the activity coefficient theoretically, we need to solve for the electric potential outside an ion, which we will model as an uniformly charged spherical shell of radius $a$. The charge distribution is given $\sigma(\mathbf{r}) = \Sigma \delta(|\mathbf{r}|-a)$, where $\Sigma=q/(4 \pi a^2)$ is the surface charge density of the ion. Substituting this charge distribution into Equation (\ref{EL-eqn}) and setting the external potential $\delta V(\mathbf{r}) =0$, we arrive at  
\begin{equation}
\nabla^2 \phi -   \frac{1}{\lambda_S^2} \phi = -\frac{4\pi}{\epsilon} \Sigma \delta(|\mathbf{r}|-a). 
\label{DH_mod}
\end{equation} 
where 
\begin{equation}
\phi(\mathbf{r})= -\frac{e}{\epsilon} \int  \frac{q \rho(\mathbf{r}') + \sigma(\mathbf{r}')}{|\mathbf{r} - \mathbf{r'}|} \; \mathrm{d} \mathbf{r'}
\end{equation} 
is the electric potential. Equation (\ref{DH_mod}) can be solved to yield 
\begin{equation}
\phi(r) =
\begin{cases}
- \frac{q}{\epsilon r} \frac{e^{a/\lambda_S}}{1+a/\lambda_S} e^{- r/\lambda_S} & a<r, \\
-\frac{q}{\epsilon a} \frac{1}{1+a/\lambda_S} & 0<r<a.
\end{cases}
\label{DH_solution}
\end{equation} 
Equation (\ref{DH_solution}) captures the physics that the self-energy of an ion is reduced by the surrounding ionic atmosphere. This reduction in self-energy due to the ionic atmosphere is given by 
\begin{equation}
\phi_{\mathrm{self}} = -\frac{q}{\epsilon a} \frac{1}{1+a/\lambda_S} + \frac{q}{\epsilon a} = - \frac{q}{\epsilon} \frac{1}{\lambda_S+a}. 
\end{equation} 
The excess chemical potential due to the ion-ion correlation can thus be computed by the Debye charging process: we consider the ionic atmosphere fixed, and compute the work done required increase the charge of the ion from $0$ to $q$ amid the ionic atmosphere
\begin{equation}
\frac{\mu_{\mathrm{ex}}}{k_B T} = \int_0^{q} \frac{\phi_{\mathrm{self}}(q')}{k_B T} \mathrm{d}q' = - \frac{1}{2} \frac{ l_B}{\lambda_S+ a}. 
\label{chemical_pot}
\end{equation} 
Therefore, the activity coefficient predicted using the semi-phenomenological model (\ref{free_energy}) is indeed the classic Debye-H\"{u}ckel expression but with the Debye length replaced by the experimentally measured screening length. 
 
\begin{figure}
\centering
\includegraphics[scale=0.45]{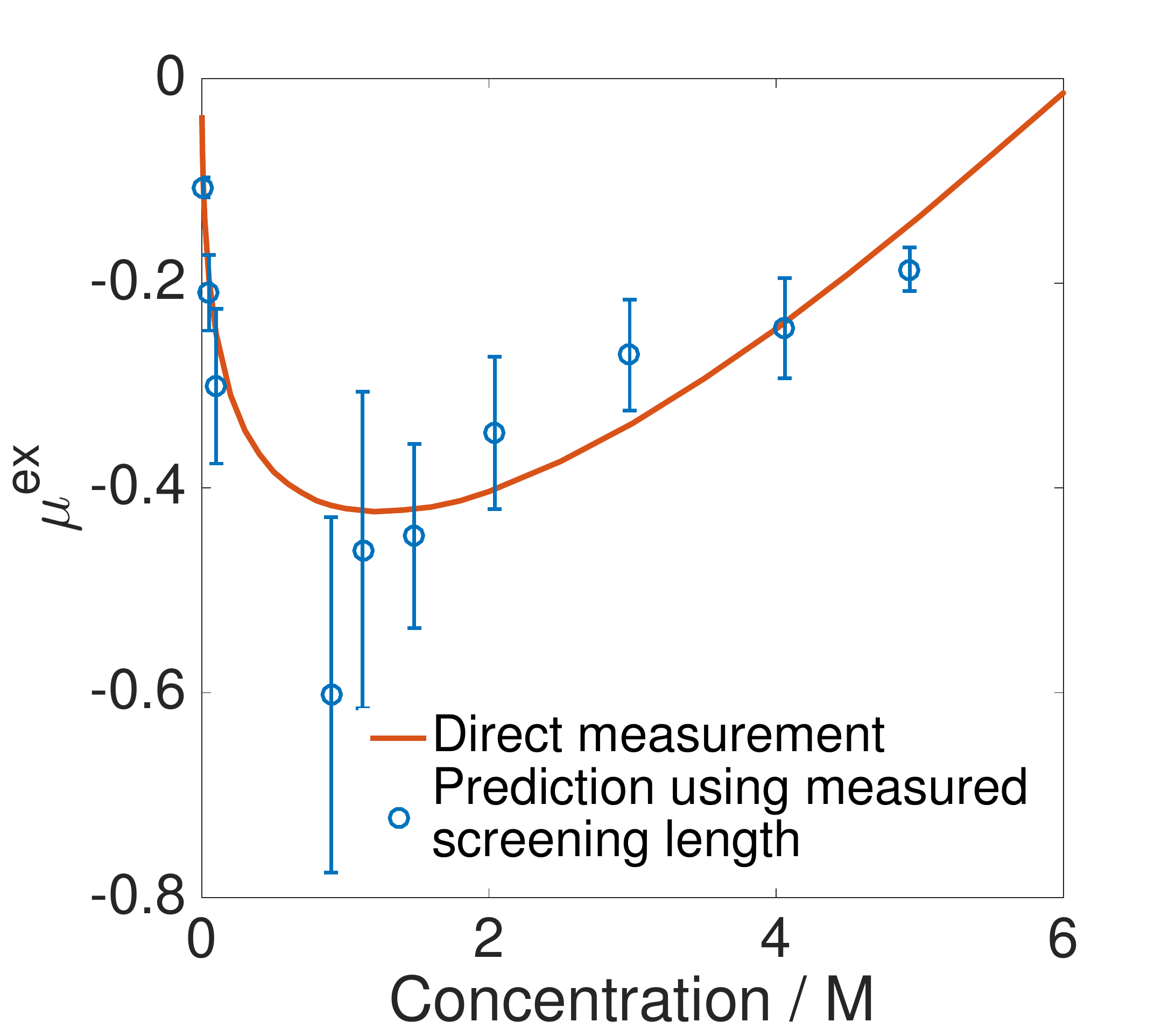} 
\caption{The activity coefficient of aqueous sodium chloride solutions predicted using Equation (\ref{chemical_pot}) and the experimentally measured screening length agrees with direct measurements \cite{hamer1972osmotic}.}
\label{activity_coeff}
\end{figure}

Figure \ref{activity_coeff} shows that the activity coefficient/excess chemical potential for aqueous sodium chloride computed using Equation (\ref{chemical_pot}) and the experimentally determined screening length agrees quantitatively with direct measurements of the activity coefficient \cite{hamer1972osmotic}. The experimental measurements of the screening length and estimates of the Bjerrum length are outlined in Section (\ref{Expt_measurements}). In particular, the increase in the experimentally measured screening length explains the upturn in the excess chemical potential which is usually attributed to excluded-volume interactions \cite{robinson2002electrolyte} or dielectric saturation \cite{valisko2014effect,shilov2015role}. 
 
\subsection{Differential capacitance at the point of zero charge}
\begin{figure}
\centering
\includegraphics[scale=0.5]{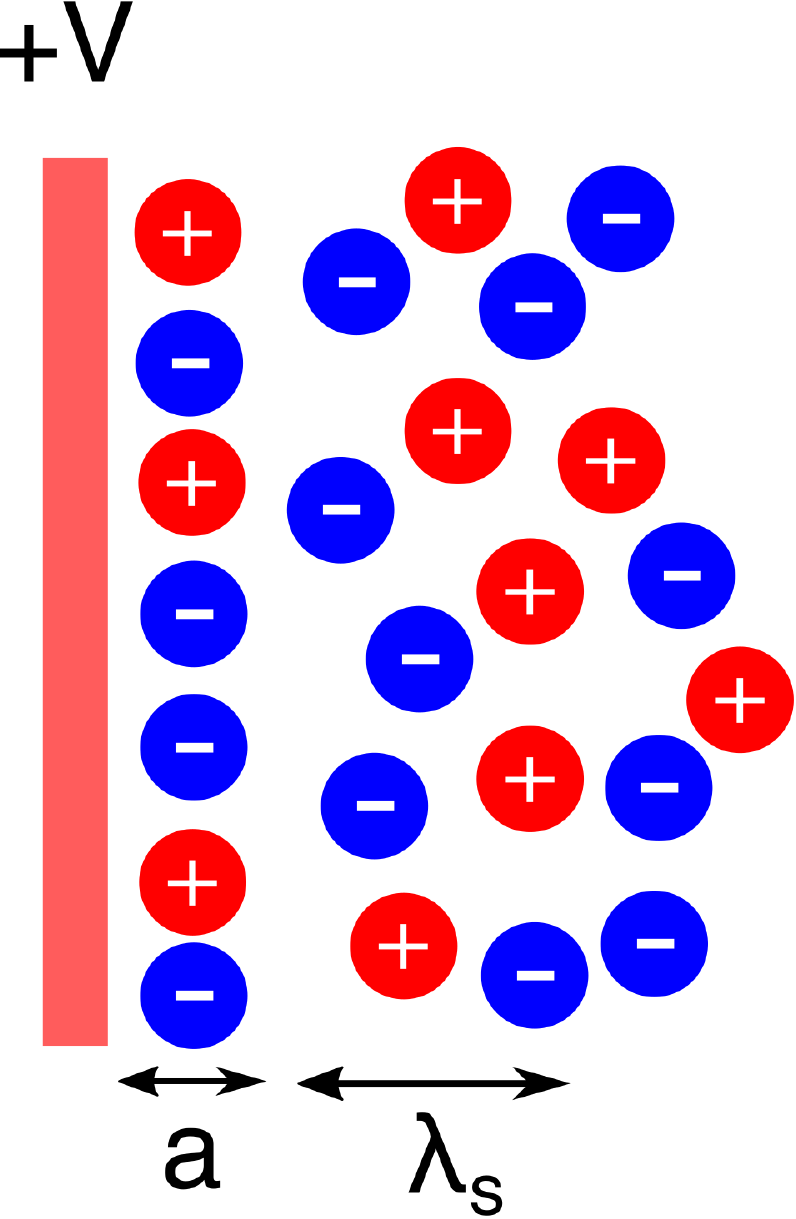}
\caption{Schematic sketch of the Gouy-Chapman-Stern model of the electrical double layer: a monolayer of ions is adsorpted on the electrode surface, and a diffuse layer loosely attached to the surface via electrostatic interactions.}
\label{GCS_model}
\end{figure}

A quantity crucial to energy storage using electrical double layer supercapacitors is the differential capacitance, defined as 
\begin{equation}
C_\mathrm{d} = \frac{\partial \sigma}{\partial V}, 
\end{equation}
where $\sigma$ is the surface charge density on the electrode and $V$ is the applied potential difference. The differential capacitance can be computed using the classic Gouy-Chapman-Stern model \cite{reddy1998modern,lockett2010differential}: we assume a layer of ions adsorpted onto the electrode -- the Stern layer -- and a ``diffuse'' layer of ions adjacent to the Stern layer which is held in place by ion-electrode electrostatic interactions (Figure \ref{GCS_model}).
The total capacitance $C_d$ is the sum of the diffuse and Stern components in series, 
\begin{equation}
\frac{1}{C_d} = \frac{1}{C_{\mathrm{Stern}}} +  \frac{1}{C_{\mathrm{Diffuse}}}. 
\label{series}
\end{equation} 
The capacitance of the Stern layer can be estimated by assuming that it is a parallel plate capacitor, with one plate being the electrode and another plate being the ions, thus 
\begin{equation}
C_{\mathrm{Stern}} = \frac{\epsilon}{4 \pi a}.
\label{stern} 
\end{equation}
 
The capacitance of the diffuse layer can be computed by minimising the semi-phenomenological free energy (\ref{free_energy}) given a fixed surface potential at the interface between the Stern layer and the diffuse layer. Analogous to the classic Debye-H\"{u}ckel theory, the electric potential away from a surface with potential $\phi_0$ is simply 
\begin{equation}
\phi = \phi_0 e^{-x/\lambda_S}
\end{equation} 
where $x$ is the direction normal to the surface and $x=0$ denotes the position of the surface, the so-called Outer Helmholtz Plane. The induced surface charge is therefore given by Gauss Law, 
\begin{equation}
4 \pi \sigma = - \epsilon \frac{\mathrm{d} \phi}{\mathrm{d} x} \Big|_{x=0} = \frac{\epsilon \phi_0}{\lambda_S}, 
\end{equation} 
thus 
\begin{equation}
C_{\mathrm{Diffuse}} = \frac{\epsilon}{4 \pi \lambda_S} 
\label{diffuse}
\end{equation} 
and substituting Equations (\ref{stern}) and (\ref{diffuse}) into (\ref{series}), we arrive at the differential capacitance at the point of zero charge 
\begin{equation}
C_d = \frac{\epsilon}{4 \pi}  \frac{1}{a + \lambda_S}.
\label{diff_cap}
\end{equation}
Beyond the point of zero charge, non-linear effects such as the finite size of ions become important \cite{bikerman1942xxxix, borukhov1997steric} and thus cannot be captured by the simple free energy (\ref{free_energy}). As such, we restrict ourselves to comparing with experimental measurement of the differential capacitance at the point of zero charge. 

To our knowledge, systematic measurement of differential capacitance as a function of ion concentration is scarce. A recent study reported the differential capacitance as a function of dilution for the ionic liquid  $\mathrm{[EMIm][NTf_2]}$ in propylene carbonate and other organic solvents on glassy carbon electrode \cite{bozym2015anomalous}. Although the screening length of $\mathrm{[EMIm][NTf_2]}$ in propylene carbonate has not be measured, we will use the measured screening length of $\mathrm{[C_4 C_1 Pyrr][NTf_2]}$ in propylene carbonate as a close proxy as the two ionic liquids share similar ion sizes and chemical functional groups. 

\begin{figure}
\includegraphics[scale=0.38]{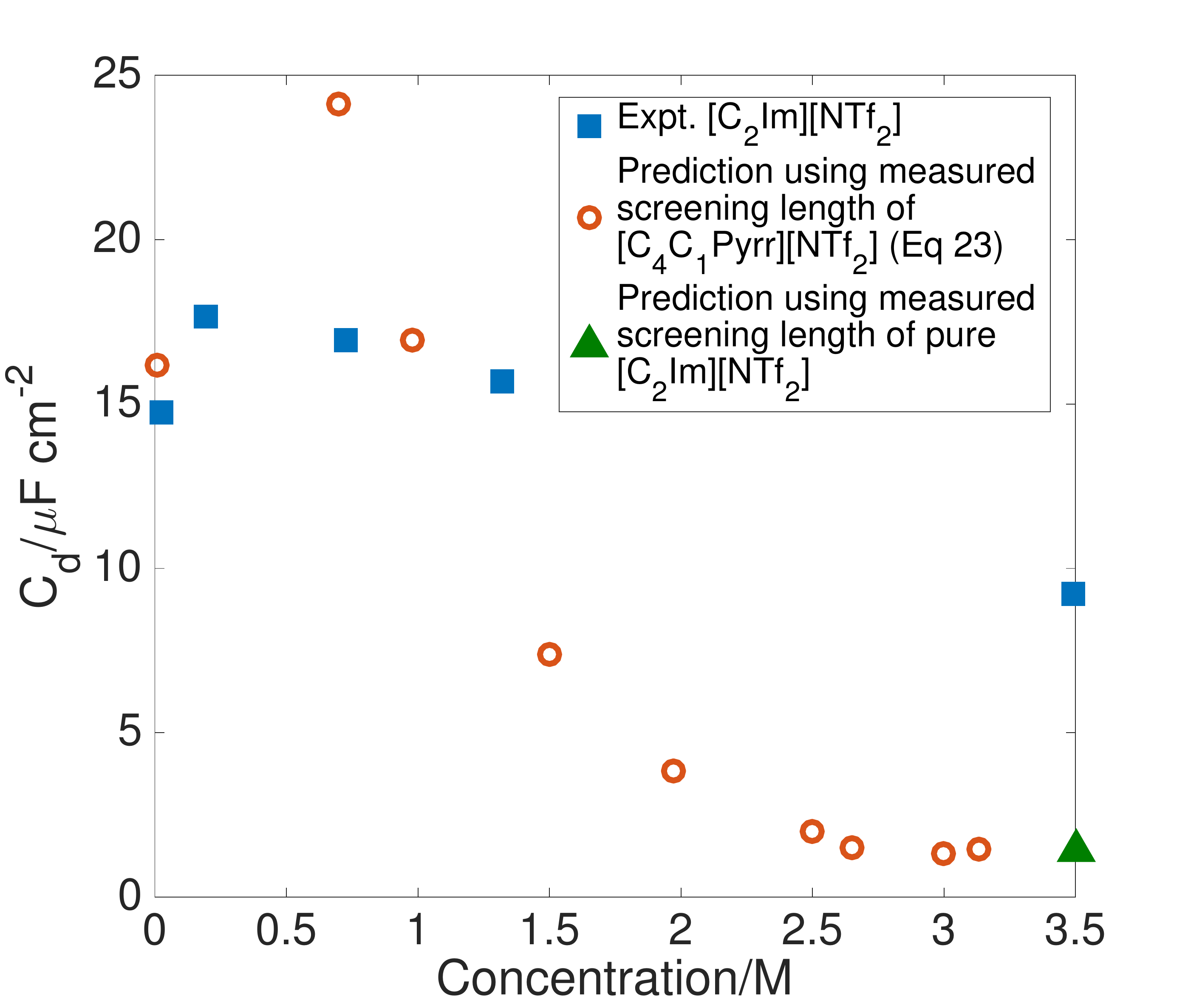}
\caption{The differential capacitance predicted by the measured screening length and Equation (\ref{diff_cap}) agrees with the observed non-monotonic dependence of the differential capacitance as a function of concentration \cite{bozym2015anomalous}. However, the predicted capacitance is significantly less than the measured capacitance as the ionic liquid-solvent mixture approaches a pure ionic liquid. The screening length of pure $\mathrm{[C_2 Im][NTf_2]}$ is taken from \cite{gebbie2015long}, and we take ion diameter $a = 4 \AA$ and dielectric constant $\epsilon = 12$ \cite{huang2011static}. }
\label{compare_capacitance}
\end{figure}

Figure \ref{compare_capacitance} shows that the differential capacitance at the potential of zero charge predicted by Equation (\ref{diff_cap}) agrees with the experimentally measured differential capacitance for low and intermediate ion concentrations. In particular, the minimum in the screening length as a function of concentration appears to match the maximum in differential capacitance -- a counterintuitive phenomenon that is outside standard Gouy-Chapman-Stern model as the classic Debye length is a decreasing function of concentration. However, the capacitance at higher concentrations cannot be captured by Equation (\ref{diff_cap}). For the pure ionic liquid, the measured capacitance is significantly larger than what one might predict from the very long screening length. Taken together, this discrepancy between screening length at high concentrations and  capacitance may indicate the dominant role of specific ion-surface interactions in determining the capacitance of ionic liquids \cite{lee2016ion}. Indeed, for pure ionic liquids, the differential capacitance is dominated by the lateral structure of the monolayer of ions nearest to the electrode \cite{limmer2015interfacial}, thus any surface chemistry or corrugations will significantly modify the differential capacitance. 

\section{Towards a theory of underscreening: a conjecture}
\label{scaling_theory}
The dependence of the screening length on the Bjerrum length and ion concentration observed empirically, Equation (\ref{underscreening}), is the opposite of the relationship that one would expect from the expression of the classic Debye length (\ref{debyehuckel}). The classic Debye length decreases with increasing ion concentration and Bjerrum length whilst the electrostatic screening length in concentrated electrolytes increases with ion concentration and Bjerrum length. In this section, we will conjecture a simple physical argument that explains this screening length.

Our argument begins with a thought experiment: Suppose we put a slab of ionic crystal between two charged surfaces, and ask whether the crystal screens the electric field. The answer is evidently no because the ions are immobile and thus the crystal acts as a dielectric slab. Now, suppose the crystal contains Schottky defects. Charge transport in such defect-laden ionic crystal occurs via ions hopping onto defect sites. Alternatively, reminiscent of the particle-hole symmetry, one could view defects itself as the charge carrier. Defects in the sub-lattice of the cations behave as negative charges, and defects in the sub-lattice of the anions behave as positive charges. Such system would be able to screen an external electric field, but the charge carrier density that enters the Debye length is the defect concentration rather than the ion concentration;  a similar conclusion is reached by analyzing the 1D lattice Coulomb fluid near close packing \cite{demery2016phase}. 

An ionic crystal is an extreme example of a correlated Coulomb melt where the ions are translationally immobile. We conjecture that a concentrated electrolyte behaves similarly to an ionic crystal in the sense that the electric potential felt by an ion due to all other ions is significantly greater than thermal fluctuations, and therefore the incentive for an ion to respond to an external potential perturbation is minimal. The role of Schottky defects is played by solvent molecules. Although solvents are charge-neutral molecules, they disrupt ion-ion correlation by freeing up a site that would have been occupied by an ion. Therefore, solvent molecules acquire an effective charge analogous to a defect in an ionic crystal. Another way to phrase the same statement is that solvent concentration fluctuations are coupled with charge fluctuations, which has been observed in molecular dynamics simulations of electrical double layer capacitors \cite{uralcan2016concentration}. 

We can put the physical intuition suggested above in a more quantitative footing by rewriting the ``defect'' Debye length   
\begin{equation} 
\Lambda_D =  (4 \pi \tilde{q}_{\mathrm{solv}}^2 l_B c_{\mathrm{solv}})^{-1/2}, 
\label{debyehuckel_defect}
\end{equation}
where $\tilde{q}_{\mathrm{solv}}^2$ is the mean-squared effective charge of a solvent molecule relative to the charge of an ion (the mean charge of a ``defect'' is zero in a symmetric electrolyte because it is as likely for a solvent molecule to be in the ``cation sub-lattice'' as in the ``anion sub-lattice''), $l_B$ is the Bjerrum length of the electrolyte, and $c_{\mathrm{solv}}$ is the concentration of solvent molecule. Assuming the system is incompressible, $c_{\mathrm{solv}}  = c_{\mathrm{tot}} - c_{\mathrm{ion}}$, where $c_{\mathrm{tot}}$ is the total concentration of the system which is assumed to be independent of ion concentration.

The next step is to estimate the effective mean-squared charge of a solvent molecule, or ``defect'', in this concentrated ionic system. Qualitatively, the defect takes the position of an ion in this correlated ionic system, and as such the energy of creating a defect must be comparable to the fluctuation energy of the ionic system per ion. The energy of a defect scales as $E_{\mathrm{defect}} \sim \tilde{q}_{\mathrm{solv}}^2$. This can be seen via symmetry (the defect energy is symmetric with respect to the charge of the defect), or by noting that a uniformly charged sphere of net charge $q$ has a self-energy that scales as $\sim q^2$. 

The energy density of the ion system can be derived using dimensional analysis: the only relevant electrostatic lengthscale in a system where Debye-H\"{u}ckel screening is negligible is the Bjerrum length. Therefore, one would expect the energy density $e_{\mathrm{ion}} \sim  l_{B}^{-3}$ from dimensional analysis. This estimate is analogous to the fluctuation energy for a dilute electrolyte which is known to scale as $\sim \lambda_D^{-3}$ \cite{safran1994statistical}, except the role of the Debye length in dilute electrolytes is replaced by the Bjerrum length in concentrated electrolytes because Debye screening is suppressed by strong ion-ion correlation. The electrostatic energy per ion is therefore  $E_{\mathrm{ion}} \sim a^3 e_{\mathrm{ele}} \sim (a/l_B)^3$. Equating  $E_{\mathrm{ion}}$ with $E_{\mathrm{defect}}$ gives the scaling relationship  
\begin{equation}
\tilde{q}_{\mathrm{solv}}^2 \sim \left(\frac{a}{l_B} \right)^3. 
\label{scale-charge}
\end{equation}
This charge scaling shows the important physics that strong ionic correlations (large Bjerrum length) suppresses thermal fluctuations in the system, and therefore the mean-squared charge of a defect which is acquired through fluctuations.   

Substituting (\ref{scale-charge}) and the incompressibility constraint into Equation (\ref{debyehuckel_defect}), we obtain 
\begin{equation}
\Lambda_D \sim  (4 \pi  ( c_{\mathrm{tot}} - c_{\mathrm{ion}})a^3/l_B^2)^{-1/2} \approx  (4 \pi c_{\mathrm{tot}}a^3/l_B^2)^{-1/2} + \frac{1}{2 \sqrt{4 \pi} (c_{\mathrm{tot}} a^3)^{3/2} }  l_B c_{\mathrm{ion}}a^3, 
\label{mod_DH}
\end{equation}  
where the expansion is valid for $c_{\mathrm{ion}} \ll c_{\mathrm{tot}} $. Equation (\ref{mod_DH}) shows that the leading order correction to Debye-H\"{u}ckel behaviour scales as $ \sim l_B c_{\mathrm{ion}}a^3$, agreeing with the scaling observed empirically ($c_{\mathrm{tot}} a^3$ in the denominator is the total packing fraction of molecular species and is approximately a constant independent of concentration). We note that for ionic liquids, although there are no solvent molecules per se, the internal degrees of freedom in the ions, in particular the alkyl chains on the cation, could perform the role of the solvent by disrupting order in the strongly correlated ionic melt. 

We next consider the ion concentration at which this ``ionic crystal'' analogy becomes appropriate. The discussion above suggests that the ionic crystal regime is reached when the typical ion-ion electrostatic interaction energy is greater than $k_B T$. We can put this inituition in a more quantitative footing: Consider a spherical blob of electrolyte of radius $R$ in the bulk electrolyte. Modelling the blob as an uniformly charged sphere, the fluctuation energy of the blob is given by 
\begin{equation} 
E_{\mathrm{fluct}} \sim k_B T l_B \frac{\left<Q^2\right>}{R}
\label{fluct_energy}
\end{equation} 
where $Q$ is the charge of the blob. If charge fluctuations in the blob follow Gaussian statistics, then $\left<Q^2\right> \sim N_{\mathrm{ion}}$ where $N_{\mathrm{ion}}$ is the number of ions in the blob, which in turn is related to the bulk density via $N_{\mathrm{ion}} \sim c_{\mathrm{ion}} R^3$. Therefore $E_{\mathrm{fluct}} \sim k_B T l_B c_{\mathrm{ion}} R^2 $ and the fluctuation energy increases with the blob size. The minimal blob size is obviously the ion diameter, and the strong correlation regime is reached when the fluctuation energy of even this minimal blob is above $k_B T$. In other words 
\begin{equation}
 l_B c_{\mathrm{ion}} a^2 \sim 1
 \label{fluct_energy_kt}  
\end{equation} 
The scaling relationship (\ref{fluct_energy_kt}) can be rewritten as $a/\lambda_D \sim 1$, which agrees with experimental results. 

We emphasise that the arguments presented above must be read as speculative conjectures. Key steps such as assuming that scaling for the fluctuation energy of the correlated ion melt has only the Bjerrum length as the relevant lengthscale and ignoring the possible dependence of the prefactor of  $E_{\mathrm{defect}}$ on the ion concentration all require more rigorous justifications. Nonetheless, we believe ideas about solvent molecules being effective charge carriers in a concentrated ionic melts may suggest that an analytical theory of asymptotically concentrated electrolytes like ionic liquids could be within reach. 

\section{Discussion and Conclusion}
We have adressed the recent demonstration of anomalously long screening lengths in concentrated electrolytes and put forward a scaling law, termed underscreening, that appears robust in experiments where solvent dielectric and electrolyte concentration are varied separately.  We hypothesise that underscreening could be seen in many more systems other than surface forces, activity coefficient and capacitance. The obvious experimental candidate is the interactions between charged colloids in concentrated electrolytes \cite{C4CP00804A}. Other candidate systems include the rate of electrochemical reaction as a function of the spectator ion concentration as the redox rate is dependent on the potential drop near the electrode, which in turn depends on the screening length. Probing the bulk correlation length using small angle scattering techniques \cite{salmon2006} or molecular simulations could reassure us the connection between asymptotic decay of surface forces and bulk properties. 

Our paper identifies several open questions, perhaps the most pressing of which is development of a rigorous theory of underscreening. We have identified two avenues towards building a microscopic model: First, the fact that our semi-phenomenological free energy (\ref{free_energy}) agrees with measured activity coefficient and to some extent differential capacitance suggests that the screening length has origin in a large local energy penalty for the accumulation of charge density. One should revisit classic theories of electrolyte solutions to identify the physics that may give rise to such local energy penalty, bearing in mind that it cannot be specific to the chemistry of the ions because this scaling is robust for a diverse class of electrolytes. Second, the scaling argument presented in Section \ref{scaling_theory} suggests that perhaps one could construct a theory of concentrated electrolyte by considering a dilute theory of interacting solvent molecules with a fluctuation-induced charge. Systematically averaging out over ions degree of freedom to arrive at a representation based on interacting ``holes'' is the analytical challenge. 
  
The physical quantities and measurements that we have mentioned thus far are equilibrium properties. The next frontier is dynamic or non-equilibrium effects. We expect that underscreening may manifest itself in linear response quantities such as conductivity, which is related to the equilibrium structure via the fluctuation-dissipation theorem. Extending the free energy (\ref{free_energy}) to understand linear non-equilibrium response and comparing with experimental data is clearly the next step. The physics beyond linear response is much richer. For example one could imagine that there is a threshold electric field above which the migration of the strongly correlated ions under the applied electric field dominates over ion-ion correlations and thus underscreening becomes unimportant; continuing the analogy between ionic crystals and concentrated electrolyte, this threshold electric field may be analogous to dielectric breakdown. Indeed, the dissociation constant of weak electrolytes is known to be an increasing function of electric field strength \cite{onsager1934deviations,kaiser2013onsager}, although a simple argument shows that underscreening cannot be understood by simple ion pairing \cite{lee2014room}. For pure ionic liquids, the fact that they comprise domains of alkyl chains and domains of charged groups with locally heterogeneous dynamics \cite{del2004structure,wang2005unique,hu2006heterogeneity,hayes2015structure} will complicate the microscopic picture of ion transport.

In summary, we have presented a series of experimental results showing that the interaction between charged surfaces in a concentrated electrolyte decays exponentially with a decay length that follows the scaling relationship $\lambda_S \sim l_B c_\mathrm{ion} a^3$, where $l_B$ is the Bjerrum length, $c_\mathrm{ion}$ the ion concentration and $a$ the ion diameter. This scaling relationship is robust to varying the chemical functionalities or molecular features of the ions, and is verified for both ionic liquid solutions and alkali halide solutions. This anomalously long screening length which increases linearly with $l_B$ and $c_\mathrm{ion}$ is the opposite of what one would expect from the classic Debye length and is termed ``underscreening''. By constructing a semi-phenomenological free energy, we show that underscreening explains the classic measurements that the activity coefficient in aqueous sodium chloride solution is a non-monotonic function of ion concentration. Underscreening also explains the observation that the differential capacitance at the point of zero charge is a non-monotonic function of ion concentration. We conjecture that in a concentrated electrolyte with strong ion-ion correlations, it is the neutral solvent molecules rather than ions that acts as charge carriers; the solvent molecules acquire an effective charge through thermal fluctuations.  We show that the empirically observed scaling relationship $\lambda_S \sim l_B c_\mathrm{ion} a^3$ follows naturally from this heuristic conjecture. 

\acknowledgments
AAL is supported by a George F. Carrier Fellowship at Harvard University. AMS is supported by a Doctoral Prize from the EPSRC. SP and CPM are supported by The Leverhulme Trust (RPG-2015-328) and the ERC (under Starting Grant LIQUISWITCH). AAL and SP thank S Safran and J Klein for their hospitality at the Weizmann Institute and insightful discussions. We are also grateful to R Evans, J Forsman, D Limmer, P Pincus, N Green, R van Roij, E Trizac and participants in the Workshop on Anomalous Screening at the Weizmann Institute for many interesting ideas and comments.

\bibliography{underscreening_refs} 
\end{document}